\documentclass[12pt,a4paper,tightenlines,nofootinbib]{revtex4}

\usepackage{amssymb}
\usepackage{epsf}

\begin{document}

\author{Ariel M\'egevand} %
\email{megevand@mdp.edu.ar} %
\altaffiliation{Member of CONICET, Argentina} %
\affiliation{Departamento de F\'{\i}sica, Facultad de Ciencias
Exactas y Naturales, Universidad Nacional de Mar del Plata, De\'an
Funes 3350, (7600) Mar del Plata, Argentina}
\author{Alejandro D. S\'anchez} %
\email{sanchez@mdp.edu.ar} %
\altaffiliation{Member of CONICET, Argentina} %
\affiliation{Departamento de F\'{\i}sica, Facultad de Ciencias
Exactas y Naturales, Universidad Nacional de Mar del Plata, De\'an
Funes 3350, (7600) Mar del Plata, Argentina}
\title{Supercooling and phase coexistence in cosmological
phase transitions} 

\begin{abstract}
Cosmological phase transitions are predicted by Particle Physics
models, and have a variety of important cosmological consequences,
which depend strongly on the dynamics of the transition. In this
work we investigate in detail the general features of the
development of a first-order phase transition.  We find
thermodynamical constraints on some quantities that determine the
dynamics, namely, the latent heat, the radiation energy density and
the false-vacuum energy density. Using a simple model with a Higgs
field, we study numerically the amount and duration of supercooling
and the subsequent reheating and phase coexistence. We analyze the
dependence of the dynamics on the different parameters of the model,
namely, the energy scale, the number of degrees of freedom and the
couplings of the scalar field with bosons and fermions. We also
inspect the implications for the cosmological outcomes of the phase
transition.
\end{abstract}

\maketitle

\section{Introduction}

Particle Physics models predict the occurrence of several phase
transitions in the early Universe, such as e.g., the electroweak
phase transition or the quark-hadron phase transition. Phase
transitions in the early Universe may leave observable vestiges,
such as topological defects \cite{vs94}, magnetic fields
\cite{gr01}, the baryon asymmetry of the Universe \cite{ckn93},
baryon inhomogeneities \cite{w84,h95}, gravitational waves
\cite{gs07} or black holes \cite{khlopov}. The effects of some of
these relics can constrain the model, as in the case, e.g., of
monopoles and domain walls. Late time cosmological phase transitions
have also been proposed to act as seeds of the large-scale structure
formation and as an explanation of the dark energy problem
\cite{fhw92,g00,chn04,m06}. The outcome of a phase transition
depends, both quantitatively and qualitatively, on several aspects
of the dynamics, for instance, the nucleation rate, the velocity of
bubble expansion, and the temperature variation during the
development of the transition.

In general, the evolution of a first-order phase transition can be
divided in three stages, namely, supercooling, reheating and phase
coexistence. At $T=T_{c}$ the free energy has two degenerate minima
separated by a barrier. Hence, the bubble nucleation rate $\Gamma$
vanishes. At $T<T_c$, ``critical'' bubbles nucleate. These are
bubbles of the true vacuum which are large enough that their volume
energy dominates over their surface tension, so they can expand.
Assuming that the standard picture of bubble nucleation applies,
bubbles of the supercooled phase will nucleate in a homogeneous
background of true vacuum. The number of bubbles will not be
appreciable until a lower temperature $T_{N}$, which can be estimated
as follows. The age of the Universe is $t\sim H^{-1}$, and a causal
volume is $V_H\sim t^3$. Then, if at least one bubble is to be
created in a time $\sim t$ in a volume $\sim V_H$, we must require
that $\Gamma t^4\gtrsim 1$. Thus, the temperature $T_N$ is roughly
determined by the condition $\Gamma\sim H^4$.

In fact, this picture may not work and the supercooling stage may be
shorter (or not occur at all). For instance, the presence of
impurities (such as e.g. topological or non-topological solitons)
could trigger bubble nucleation \cite{impur}. Also, if the phase
transition is weakly first-order, i.e., if the barrier of the free
energy is sufficiently small, thermal fluctuations called subcritical
bubbles may dominate \cite{gkw91}. In this case, there may be a
two-phase emulsion already at $T=T_c$. Then, sub-critical bubbles may
percolate and true-vacuum domains may begin to grow at a temperature
$T>T_N$.

Initially, bubbles of true vacuum grow with a velocity which is
governed by the pressure difference across their walls and by the
viscosity of the hot plasma or relativistic gas surrounding them. As
bubbles expand, latent heat is liberated and reheats the system back
to a temperature $T_r$. As a consequence, the expansion of bubbles
slows down, since the pressure difference decreases as $T$ approaches
$T_c$. If the latent heat $L$ is negligible, there will be no
temperature variation. One expects that reheating will be important
if $L$ provides the energy density difference needed to increase the
temperature of radiation from $T_{N}$ back to $T_r\approx T_{c}$,
i.e., when $L\sim \delta \rho _{R}\sim T_{c}^{4}-T_{N}^{4}$. If $L$
is much larger than $\delta \rho _{R}$, the temperature $T_r$ will be
very close to $T_{c}$. When this happens,  a stage of ``slow growth''
or ``phase coexistence'' follows. Indeed, since $T$ cannot increase
beyond $T_c$, bubbles will grow only at the rate at which the
expansion of the Universe takes away the injected energy. The
temperature will thus remain nearly constant until every region of
space has been converted to the stable phase.

Although the above picture is quite general, the details of the
dynamics depend on the specific model. A complete analysis involves,
even in the simplest cases, solving a set of integro-differential
equations for the nucleation and expansion of bubbles, which takes
into account the reheating of the thermal bath. Therefore, it is
useful to find general characteristics, which will permit to obtain
some conclusions before embarking on the task of computing the
development of a given phase transition. In Ref. \cite{m04}, an
analytical approach was performed, which allowed to obtain some
general conclusions on the evolution. However, due to the involved
dynamics of reheating, the analytical study requires some rough
approximations, particularly for the nucleation rate. A numerical
investigation is thus necessary in order to have a better
understanding of the dynamics of first-order phase transitions and
their cosmological consequences.

In this work, we shall perform a detailed study of the general
dynamics of phase transitions. We shall be interested in first-order
phase transitions occurring either in the radiation dominated epoch,
or in a sector composed of radiation.  In particular, we shall
examine thermodynamic constraints which apply to any first-order
phase transition. As we shall see, this allows to discuss on the
possible effects of a model without making numerical calculations.
We shall also make a numerical investigation of the dynamics. For
that purpose, we shall use a simple model for the free energy, which
allows to consider different kinds of phase transitions, both weak
and strong. The model also provides an approximation for realistic
theories (e.g., different extensions of the Standard Model). We
shall discuss the implications of our results for the cosmological
outcomes of the phase transition.

The article is organized as follows. In the next section we discuss
some general properties of phase transition dynamics, and we study
model-independent relations between thermodynamical parameters.
Then, in section \ref{themodel} we consider a simple model,
consisting of a scalar (Higgs) field, which has Yukawa couplings to
different species of bosons and fermions. We write down the one-loop
finite-temperature effective potential for this model, and discuss
the different kinds of phase transitions the model can present. In
section \ref{transi} we consider the equations for the evolution of
the phase transition, and we compute them numerically. We are
particularly concerned with the amount and duration of supercooling,
and with the extent of the phase coexistence stage.

We apply the results of this investigation in section \ref{cosmo},
where we analyze some of the possible cosmological outcomes of a
phase transition to illustrate the effect of the dynamics. We
consider the formation of baryon inhomogeneities in the electroweak
phase transition, the creation of topological defects, and the
generation of magnetic fields. We also discuss on different
proposals of late-time phase transitions as solutions to the
dark-energy problem. We show that thermodynamical constraints rule
out some of these models. Our conclusions are summarized in section
\ref{discus}. Some technical details of the calculation are left to
the appendix.

\section{Phase transition and thermodynamic parameters \label{termo}}

We can use thermodynamic considerations to obtain some general
information on the amounts of supercooling and reheating and on the
duration of the phase transition, without specifying the form of the
free energy.

\subsection{Supercooling and phase coexistence}

Consider a system which undergoes a phase transition at a
temperature $T_c$. The high-temperature phase consists only of
radiation and false vacuum energy, so the energy density is of the
form
\begin{equation}
\rho _{+}=\rho _{\Lambda }+\rho _{R},  \label{rho+}
\end{equation}%
where $\rho _{\Lambda }$ is a constant and $\rho _{R}=g_*\pi
^{2}T^{4}/30$, where $g_*$ is the number of relativistic degrees of
freedom (d.o.f.). At the critical temperature $T_{c}$ the two phases
have the same free energy density, but different energy density. The
discontinuity  $\Delta\rho (T_{c})\equiv\rho _{+}(T_{c})-\rho
_{-}(T_{c})$, with $\rho _{-}$ the energy density of the
low-temperature phase, gives the latent heat
\begin{equation}
L\equiv\Delta\rho (T_{c})=T_{c}\Delta s(T_{c}),  \label{lat}
\end{equation}%
with $\Delta s=s_{+}-s_{-}$ the entropy density difference. This
entropy is liberated as regions which are in the high-$T$ phase
convert to the low-$T$ one.

Since entropy is conserved in the adiabatic expansion of the
Universe, the entropy density of the system can be written as
\begin{equation}
s=s_{+}(T_{c})(a_{i}/a)^{3},  \label{entropy}
\end{equation}%
where $a$ is the scale factor, and $a_{i}$ is its value at the
beginning of the transition, i.e., at $T=T_{c}$. During the phase
transition, $s$ is given by
\begin{equation}
s=s_{+}(T)-\Delta s(T)f,  \label{entroptrans}
\end{equation}%
where $f$ is the fraction of volume occupied by bubbles of low-$T$
phase.

If there is little supercooling (e.g., if the phase transition is
weakly first-order, or if bubble nucleation is triggered by
impurities), the temperature $T_N$ at which bubbles form and start
to grow will be very close to $T_{c}$. In this case, a small $L$ can
take the system back to $T_{c}$. Then, a good approximation is to
consider that the phase transition develops entirely at $T=T_{c}$,
with equilibrium of phases \cite{w84,s82}. Thus, the fraction of
volume is easily obtained from Eqs. (\ref{entropy}) and
(\ref{entroptrans}). The result is \cite{m04}
\begin{equation}
f=\frac{s_{+}(T_{c})}{\Delta s(T_{c})}\left[
1-\left(\frac{a_{i}}{a}\right)^{3}\right]. \label{fcoex}
\end{equation}
The phase transition completes when $f=1,$ so its duration is
determined by the condition
\begin{equation}
\left( a_{i}/a_{f}\right) ^{3}=1-\Delta s(T_{c})/s_{+}(T_{c}),
\label{af}
\end{equation}%
where $a_{f}$ is the scale factor at the end of the phase
transition.

In general, though, bubble nucleation does not begin as soon as $T$
reaches $T_{c}$. The temperature decreases until the nucleation rate
becomes comparable to the expansion rate. During supercooling, the
entropy of the system is that of radiation, $s_{+}(T)=s_R(T)$, with
\begin{equation}
s_{R}(T)=\frac{4}{3}\frac{\rho _{R}}{T} =\frac{2g_*\pi
^{2}}{45}T^{3}, \label{s+}
\end{equation}%
so, from Eq. (\ref{entropy}) we have $T=T_{c}a_{i}/a$. When the
number of bubbles becomes noticeable, the released entropy begins to
reheat the system. The minimum temperature $T_{m}$ delimits the end
of supercooling. It is reached at a value $a_m$ of the scale factor
given by $ T_{m}\approx T_{c}a_{i}/a_{m}.$ One expects that for
$L\gtrsim \delta \rho _{R}\equiv \rho _{R}(T_{c})-\rho _{R}(T_{m})$,
the temperature will go back to $T\approx T_{c}$ and a period of
phase coexistence will begin. We will now show that the condition
for phase coexistence to occur is in fact
\begin{equation}
\Delta s(T_{c})> \delta s_{R}, 
\label{condphcoex}
\end{equation}%
where $\delta s_{R}\equiv s_{R}(T_{c})-s_{R}(T_{m})$. In terms of
energy, we have $T_{c}\Delta s(T_{c})=L$ and  $\delta s_{R} = (4/3)
\delta \left(\rho _{R}/T \right)$, so the above condition becomes
$L\gtrsim \left( 4/3\right) \delta \rho _{R}$.

Assuming that a phase coexistence stage at $T_r\approx T_c$ is
reached, we can go back to Eqs. (\ref{entropy}) and
(\ref{entroptrans}), which lead again to the result (\ref{af}) for
the {\em total} change of scale $a_f/a_i$, even though this time the
temperature was not constant from the beginning. Therefore, the final
value of the scale factor $a_{f}$ is not affected by the previous
supercooling and reheating stages. This will only be possible,
however, if $a_{m}<a_{f}$, since the supercooling stage cannot be
longer than the total duration of the phase transition. During
supercooling, $s=s_+$, so $s_+(T_m)$ is given by Eq. (\ref{entropy})
with $a=a_m$. Comparing with Eq. (\ref{af}), the condition
$a_{m}<a_{f}$ gives $s_{+}(T_{m})> s_{+}(T_{c})-\Delta s(T_{c})$.
Since $s_+=s_R$,  Eq. (\ref{condphcoex}) follows.

The value of $L$ can be easily calculated for any model, since it is
derived directly from the free energy. In contrast, calculating
$\delta \rho _{+}$ entails the evaluation of the nucleation rate
$\Gamma $, which must be calculated numerically, and then solving
the equations for the evolution of the phase transition in order to
determine $T_{m}$. We will perform such calculation in section
\ref{transi}. Provided that condition (\ref{condphcoex}) is
fulfilled, the value of $a_{f}$ will be independent of the amount of
supercooling, and given  by Eq. (\ref{af}). We can write
equivalently
\begin{equation}
\left( a_{i}/a_{f}\right) ^{3}=1-3L/4\rho _{R}. \label{af2}
\end{equation}
How long will the phase transition go on, depends on how large $L$
is. Since the entropy difference is bounded by $\Delta
s(T_{c})<s_{+}(T_{c})$, the latent heat has a maximum value $ L_{\max
}=T_{c}s_{+}=4\rho _{R}/3$. We see that $a_{f}\to\infty $ in this
limit. This is because $s_{-}=0$, so all the entropy must be
extracted from the system in order to complete the phase transition,
and this requires an infinite amount of work.

The duration $\Delta t$ of the phase transition is related to the
expansion factor $a_{f}/a_{i}$ through the expansion rate $H$.
Consequently, it depends on the different kinds of energy (e.g.,
matter, vacuum, radiation) that make up the total energy density
$\rho $. If our system is uncoupled from other sectors (as in the
case of late-time phase transitions), then it is not straightforward
to calculate $\Delta t$. In the early Universe, instead, we can
assume that all particle species are in equilibrium with each other
and constitute a single system which is dominated by radiation.
Then, for the period of phase coexistence at $T=T_{c}$, the equation
of state is especially simple, since temperature and pressure are
constant. The energy density is given by
\begin{equation}
\rho =T_{c}s_{+}\left( a_{i}/a\right) ^{3}-p_{c},  \label{rhocoex}
\end{equation}%
where%
\begin{equation}
p_{c}=\rho _{R}(T_c)/3-\rho _{\Lambda } \label{pc}
\end{equation}%
is the pressure at $T=T_{c}$. Consequently, the Friedmann
equation\footnote{We neglected a term $k/a^{2}$ in Eq.
(\ref{friedmann}). This is correct for most of the history of the
Universe.}
\begin{equation}
H^{2}\equiv \left( \frac{\dot{a}}{a}\right) ^{2}=\frac{8\pi G}{3}\rho,
\label{friedmann}
\end{equation}%
where  $G$ is Newton's constant, can be solved analytically
\cite{s82,iks86,m04}.  We have
\begin{equation}
\left( \frac{a}{a_{i}}\right) ^{3}=\frac{T_{c}s_{+}}{p_{c}}\sin ^{2}\left(
\omega \left( t-t_{i}\right) +\delta \right) ,  \label{at}
\end{equation}%
where $\omega =\sqrt{6\pi Gp_{c}}$ and $\delta =\arcsin
\sqrt{p_{c}/T_{c}s_{+}}$.

From Eqs. (\ref{af}) and (\ref{at}) we obtain
\begin{equation}
\frac{\Delta t}{\tilde{t}}=\frac{4}{3}\sqrt{\frac{\rho _{+}}{p_{c}}}\arcsin %
\left[ \frac{3/4}{\sqrt{1-\Delta s/s_{+}}}\sqrt{\frac{p_{c}}{\rho _{R}}}%
\frac{\sqrt{\rho _{+}}-\sqrt{\rho _{-}}}{\sqrt{\rho _{R}}}\right] ,
\label{deltat}
\end{equation}%
where $\tilde{t}=\left( 2H_{i}\right) ^{-1}\approx t_{i}.$ Notice that $%
\Delta t/\tilde{t}$ depends only on the two parameters $r=L/\rho_R$
(equivalently, $\Delta s/s_{+}$) and $R=\rho _{\Lambda }/\rho _{R}$.
We remark that, as long as a temperature $T_r\approx T_{c}$ is
reached after reheating, $\Delta t$ gives the {\em total} duration
of the phase transition, i.e., the time elapsed from the beginning
of supercooling at $t=t_{i}$ until the end of phase coexistence at
$t=t_{f}$. As we have seen, the condition for the validity of Eq.
(\ref{deltat}) is that supercooling ends before this time.
Otherwise, $\Delta t$ will be given essentially by the duration of
supercooling, since the subsequent reheating stage will be short. In
that case, Eq. (\ref{deltat}) gives just a lower bound for the
duration of the phase transition.

\subsection{Constraints on thermodynamic parameters}

At the critical temperature, one expects that the energy density of
radiation is at least of the order of that of the false vacuum,
since radiation must provide the entropy necessary to make the
minima of the free energy degenerate. Notice that the exact relation
between $\rho_R(T_c)$ and $\rho_{\Lambda}$ can be determinant for
the dynamics of phase coexistence. Indeed, for $\rho _{R}/3<\rho
_{\Lambda }$, the pressure $p_{c}$ is negative and the sine in Eq.
(\ref{at}) becomes a hyperbolic sine, which indicates that the
expansion of the Universe is accelerated. This happens because the
energy density (\ref{rhocoex}) includes a constant term $\rho
_{\Lambda }^{\mathrm{eff}}=-p_{c}$, which represents an effective
cosmological constant \cite{s81,m06}. In this case,
$\rho_{\Lambda}^{\rm eff}>0$. Moreover, if $\rho _{\Lambda }\sim
\rho _{R}(T_c) $, the false vacuum energy may become important
before the phase transition, i.e., at $T\gtrsim T_c$.

On the other hand,  if $\rho _{\Lambda }<\rho _{R}/3$, we have $\rho
_{\Lambda }^{\mathrm{eff}}=-p_{c}<0$. Then, according to Eq.
(\ref{at}) the Universe will collapse after a time $t_{c}\sim
1/\omega $, unless phase coexistence ends before this time, so that
this equation is no longer valid. Notice that phase coexistence may
be long if $L\approx 4\rho_R/3$. The collapse occurs because the
energy density (\ref{rhocoex}) and, consequently, the expansion rate
(\ref{friedmann}) vanish for a finite value of $a/a_i$.
Nevertheless, the quantities $L$, $\rho _{R}$, and $\rho _{\Lambda
}$ are constrained by thermodynamical relations, and we will show
that none of the above situations can arise, i.e., phase coexistence
will not cause either accelerated expansion nor collapse of the
Universe.

The pressure of the relativistic system is given by
$p=-\mathcal{F}$, where $\mathcal{F}$ is the free energy density.
Hence, at $T=T_c$ we have $p_{c}=-\mathcal{F}_{+}\left( T_{c}\right)
=- \mathcal{F}_{-}\left( T_{c}\right) $. The free energy density
depends only on temperature, $d\mathcal{F}=-sdT-\left(
p+\mathcal{F}\right) dV/V=-sdT$. Since $s>0$, $\mathcal{F}\left(
T\right) $ must be a monotonically decreasing function. Therefore we
have in particular $\mathcal{F}_{-}\left( T\right)
<\mathcal{F}_{-}\left( T=0\right) $ for any $T>0$. But at $T=0$ the
free energy matches the energy. Hence, assuming that the energy
density vanishes in the true vacuum, we have $\mathcal{F}_{-}\left(
T=0\right) =\rho _{-}(T=0)=0$. Then, $\mathcal{F}_{-}\left(
T_{c}\right) <0$ and $p_{c}>0$, so the condition for accelerated
expansion is never fulfilled. Moreover, the condition
\begin{equation}
\rho _{\Lambda }<\rho _{R}(T_c)/3 \label{thconstr1}
\end{equation}
implies that false vacuum energy never dominates, unless the system
departs from thermal equilibrium (for instance, $\rho _{\Lambda }$
may become dominating in the course of supercooling).

Now, since $p_{c}>0$, the Universe will not collapse only if phase
coexistence ends before $\rho$ vanishes. According to Eq.
(\ref{rhocoex}), this is true if $\left( a_f/a_{i}\right)
^{3}<T_{c}s_{+}/p_{c}$. Using Eqs. (\ref{af2}), (\ref{s+}) and
(\ref{pc}) the condition becomes
\begin{equation}
L<\rho _{+}=\rho _{\Lambda }+\rho _{R}(T_c). \label{thconstr2}
\end{equation}
But this is always fulfilled, since $L=\rho _{+}-\rho _{-} $, and
$\rho _{-}(T)>0$ at $T>0$ [because $d\rho/dT=Tds/dT>0$ and $\rho
_{-}(T=0)=0$].

The inequalities above become equalities only for $\mathcal{F}
_{-}(T)=\rho _{-}(T)=0$, i.e., at $T=0$. So, both limiting values
$\rho _{\Lambda }=\rho _{R}/3$ and $L=\rho _{\Lambda }+\rho _{R}$
are attained only if $T_{c}= 0$. In this limit $\rho _{\Lambda }$
and $\rho _{R}$ vanish, but still  $L/\rho _{R}\to 4/3$. Hence, Eq.
(\ref{af2}) implies that $a_f\to\infty$. Thus, for a phase
transition with $T_{c}\approx 0$ we will have a long
phase-coexistence stage. For a given model with a fixed energy scale
$v,$ small $T_c$ means $T_{c}\ll v$, i.e., the metastable minimum
and the barrier must persist at $T\ll v.$ At such low temperatures,
the free energy coincides approximately with the zero-temperature
potential, and the minimum $\phi _{c}$ tends to the zero-temperature
value $v$. This corresponds to a very strongly first-order phase
transition, with $\phi _{c}/T_c\gg 1$. In this case one expects that
the nucleation rate will be suppressed and the supercooling stage
will be long too. However, it is not straightforward to compare the
duration $\Delta t_s$ of supercooling to that of phase coexistence,
since the latter depends significantly on the total number of d.o.f.
$g_*$, while $\Delta t_s$  depends essentially on the bubble
nucleation rate $\Gamma$. In section \ref{transi} we will see that,
depending on the model, we can have either $\Delta t_s\ll\Delta t$
(i.e., little supercooling) or $\Delta t_s \approx\Delta t$ (i.e.,
short phase coexistence).

In a specific model, the parameters  $\rho _{\Lambda }$, $L$, and
$\rho _{R}$ can be derived from the free energy. The constraints
(\ref{thconstr1},\ref{thconstr2}), i.e., $R\leq 1/3$ and $r\leq
R+1$, should then be automatically fulfilled\footnote{Notice that
some approximations for the free energy may allow values that fall
outside this region (see e.g. the discussion on dark-energy models
in section \ref{cosmo}).}. In general, $\rho _{\Lambda }$ and $L$
will be even more constrained. For instance, the radiation density
$\rho _{R}$ may contain a component $\rho _{l}$ from particles which
are in thermal equilibrium with the system, but are not directly
coupled to the order parameter, and therefore do not contribute to
$L$ and $\rho _{\Lambda }$ (e.g., ``light" particles which do not
acquire masses through the Higgs mechanism). The inequalities above
hold for the radiation of the system alone, i.e., $\rho _{R}-\rho
_{l},$ so the constraints become $\rho _{\Lambda }\leq \left( \rho
_{R}-\rho _{l}\right) /3$ and $L\leq \rho _{\Lambda }+\rho _{R}-\rho
_{l}$ . If $\rho_l=g_l\pi^2T^4/30$, we have $R\leq x /3$ and $r\leq
R+x$, where $x=1-g_{l}/g_*$.

\begin{figure}[htb]
\centering \epsfysize=6cm \leavevmode \epsfbox{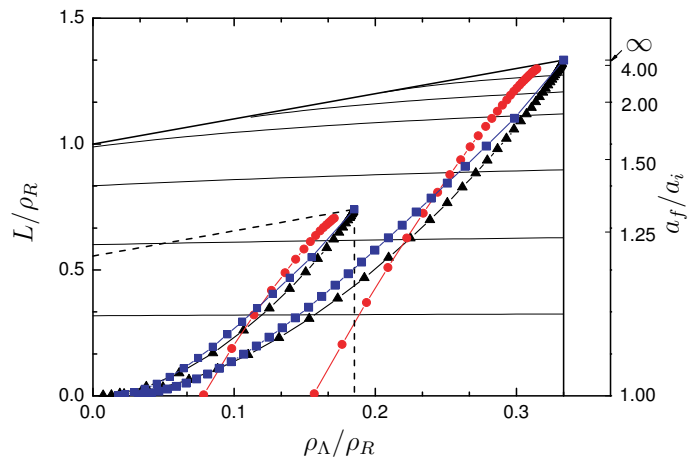}
\caption{Contours of constant time in the allowed region of the
plane $(\rho _{\Lambda }/\rho _{R}$,$L/\rho _{R})$. From bottom to
top, the curves correspond to $\Delta t/\tilde{t}=0.2,0.5,1,2,3$ and
$5$.  The points correspond to varying $h_b$ in the model of section
\ref{themodel} for $h_f=0.7$ (blue squares), $h_f=h_b$ (black
triangles), and $h_f=h_b$ with $\mu_b\neq 0$ (red circles). The
three curves on the right correspond to $g_l=0$, and those on the
left to $g_l/g_*\approx 0.44$.} \label{fconstr}
\end{figure}

Fig. \ref{fconstr} shows the region in the $(R,r)$-plane allowed by
thermodynamics, and inside that, the contours of constant time
$\Delta t$. On the right axis we have indicated some values of
$a_f/a_i$ (which depend only on $r$). The points correspond to some
of the phase transitions considered in the next section. We have
plotted two sets of curves, corresponding to $g_l=0$ and
$g_l/g_*\approx 0.44$. The dashed line delimits the allowed region
for the latter case. As the phase transition becomes stronger, the
latent heat increases. However, the limit $L=\rho_{\Lambda}+\rho_R$
is reached for $T_c\to 0$, together with the limit
$\rho_{\Lambda}=\rho_R/3$. That is why all the curves approach the
upper-right corner of the allowed region.

The analytic approximation given by  Eq. (\ref{deltat}) for the
total duration of the phase transition is valid only if condition
(\ref{condphcoex}) is satisfied. Furthermore, we cannot describe,
within this approach, the transition between  supercooling and phase
coexistence, i.e., the reheating stage. A complete description of
phase transition dynamics involves the computation of the nucleation
rate. This requires specifying a model for the free energy.

\section{The free energy} \label{themodel}

We will consider a theory described by a scalar field $\phi $ with
tree-level potential
\begin{equation}
V_{0}\left( \phi \right) =-\frac{\lambda v^{2}}{2}\phi ^{2}+\frac{%
\lambda}{4}\phi ^{4},
\end{equation}%
which has a maximum at $\phi =0$ and a minimum at $\phi =v.$  The
one-loop effective potential is of the form
\begin{equation}
V\left( \phi \right) =V_{0}\left( \phi \right) +V_{1}\left( \phi \right)
+\rho _{\Lambda },
\end{equation}%
where $V_{1}\left( \phi \right)$ is the one-loop zero-temperature
correction, and we have added a constant $\rho _{\Lambda }$ so that
the energy density vanishes in the true vacuum.  Imposing the
renormalization conditions that the minimum of the potential and the
mass of $\phi $ do not change with respect to their tree-level values
\cite{ah92}, the one-loop correction is given by
\begin{equation}
V_{1}\left( \phi \right) =\sum_{i}\pm \frac{g_{i}}{64\pi ^{2}}\,\left[
m_{i}^{4}(\phi )\left( \log \left( \frac{m_{i}^{2}(\phi )}{m_{i}^{2}(v)}%
\right) -\frac{3}{2}\right) +2m_{i}^{2}(\phi )m_{i}^{2}(v)\right] ,
\end{equation}%
where $g_{i}$ is the number of d.o.f. of each particle species,
$m_{i}\left( \phi \right) $ is  the $\phi $-dependent mass, and the
upper and lower signs correspond to bosons and fermions,
respectively.

The free energy density results from adding finite-temperature
corrections to the effective potential,
\begin{equation}
\mathcal{F}(\phi ,T)=V\left( \phi \right) +\mathcal{F}_{1}(\phi ,T),
\end{equation}%
where the one-loop contribution is
\begin{equation}
\mathcal{F}_{1}(\phi ,T)=\sum_{i}\frac{g_{i} T^{4}}{2\pi ^{2}}I_{\mp
}\left[ \frac{m_{i}\left( \phi \right) }{T}\right] , \label{f1loop}
\end{equation}%
and $I_{-}$, $I_{+}$ stand for the contributions from bosons and
fermions, respectively,
\begin{equation}
I_{\mp }\left( x\right) =\pm \int_{0}^{\infty }dy\,y^{2}\log \left( 1\mp e^{-%
\sqrt{y^{2}+x^{2}}}\right) .  \label{integrals}
\end{equation}%
For simplicity, we will consider in general masses of the form
$m_i\left( \phi \right) =h_i\phi $, where $h_i$ is the Yukawa
coupling. Thus, the free energy takes the form
\begin{eqnarray}
\mathcal{F}\left( \phi ,T\right) &=& V_0
(\phi) +\sum \frac{\pm g_ih_i^{4}}{%
64\pi ^{2}}\,\left[ \phi ^{4}\left( \log \frac{\phi ^{2}}{v^{2}}-\frac{3}{2}%
\right) +2v^{2}\phi ^{2}\right]  \nonumber \label{ffinal} \\
&& + \rho_{\Lambda}+\sum\frac{g_iT^{4}}{2\pi ^{2}}I_{\mp }\left( \frac{h_i\phi }{T}%
\right)-\frac{\pi ^{2}}{90}%
g_{l}T^{4},
\end{eqnarray}%
where the last term accounts for the contribution of species with
$h_i=0$, so $g_l$ is the effective number of d.o.f. of relativistic
particles. The constant $\rho_{\Lambda}$ is obtained by imposing
that $V\left( v\right) =0$, so
\begin{equation}%
\rho_{\Lambda}=\left(\lambda+\frac{\sum\mp g_ih_i^{4}}{32\pi
^{2}}\right)\frac{v^{4}}{4}. \label{rholambda}
\end{equation}
Notice that $\rho _{\Lambda }$ gives the energy density of the false
vacuum,  $\rho_{\Lambda }=V\left( 0\right) .$

At high temperature the free energy (\ref{ffinal}) has a single
minimum at $\phi =0$.  As the temperature decreases, a non-zero
local minimum $\phi _{m}\left( T\right) $ develops. Therefore, the
free energy in the high- and low-temperature phases is given by
$\mathcal{F}_{+}\left( T\right) \equiv \mathcal{F}\left( 0,T\right)
$ and $ \mathcal{F}_{-}\left( T\right) \equiv \mathcal{F}\left( \phi
_{m}\left( T\right) ,T\right) $, respectively. In the phase with
$\phi=0$, all particles are massless and
\begin{equation}
{\cal F}_+ = -g_*\pi^2T^4/90+\rho_{\Lambda}, \label{fmas}
\end{equation}
where $g_*=\sum g_b+(7/8)\sum g_f$ is the effective number of d.o.f.
($b$ stands for bosons and $f$ for fermions). Thus we have radiation
and false vacuum. At the critical temperature $T_{c}$, the two
minima $\phi =0$ and $\phi _{m}\left( T_{c}\right) \equiv \phi _{c}$
have the same free energy. Below this temperature, $\phi _{m}\left(
T\right) $ becomes the global minimum. In general, as temperature
decreases further the barrier between minima disappears and the
minimum at $\phi =0$ becomes a maximum. This happens at a
temperature $T_{0}$ given by
\begin{equation}
T_{0}^{2}=\frac{ \lambda + \sum \mp g_{i}h_{i}^{4}/16\pi ^{2}}{\sum
g_{b}h_{b}^{2}/12+\sum g_{f}h_{f}^{2} /24} v^{2}. \label{t0}
\end{equation}%
Finally, at zero temperature we have ${\cal F}(\phi,0)=V(\phi)$, so
$\phi _{m}(0) =v$. Notice, however, that the {\em zero-temperature}
boson contribution may turn the maximum at $\phi=0$ of the
tree-level potential $V_0(\phi)$ into a minimum of $V(\phi)$.  In
this case, there will be two minima still at $T=0$. Indeed, for
$\sum g_{b}h_{b}^{4}\geq \sum g_{f}h_{f}^{4}+ 16\pi ^{2}\lambda$,
the r.h.s. of Eq. (\ref{t0}) becomes negative, which means that the
barrier never disappears. Furthermore, for strongly coupled bosons
the origin can become the stable zero-temperature minimum. Indeed,
for $\sum g_{b}h_{b}^{4}\geq \sum g_{f}h_{f}^{4}+ 32\pi ^{2}\lambda
$ the vacuum energy density (\ref{rholambda}) becomes negative. In
that case, the origin is stable at all temperatures, and there is no
phase transition.

The energy density can be derived from the free energy by means of the
relations $\rho=Ts+{\cal F}$ and $s=-d\mathcal{F}/dT$. Thus, from
Eq. (\ref{fmas}) we obtain $\rho_+=\rho_{\Lambda}+\rho_R$, and
$\rho_-=-T{\cal F}'_-+{\cal F}_-$. At $T=T_{c}$, $\mathcal{F}_{+}=
\mathcal{F}_{-}$, so the latent heat is $L=-T_{c}\Delta {\cal F}'$.
Taking
into account that $\partial \mathcal{F}/\partial \phi |_{\phi =\phi _{m}}=0$,
we find
\begin{equation}
L=\sum \frac{2g_{i}T_c^{4}}{\pi ^{2}}\left[ -I_{\mp }\left( 0\right)
+I_{\mp }\left( \frac{h_i\phi _{c}}{T_{c}}\right) -\frac{h_i\phi
_{c}}{4T_{c}}I_{\mp }^{\prime }\left( \frac{h_i\phi
_{c}}{T_{c}}\right) \right] . \label{lat1}
\end{equation}%
The functions $I_{\pm }\left( x\right) $ are negative and
monotonically increasing, so we see that the one-loop effective
potential satisfies the thermodynamical bound $L\leq\sum
-2g_{i}T_c^{4}I_{\mp }\left( 0\right) /\pi ^{2} = 4\rho _{R}/3$.
Furthermore, $I_+$ and $I_-$ fall exponentially for large $x$.
Therefore, $L$ approaches the limit $L/ \rho _{R}\to 4/3$ for
$h_i\phi_c/T_c\to \infty$.

For our purposes it will be sufficient to consider only four
particle species, namely, two bosons and two fermions. In this way
we can have weakly coupled fermions and bosons with Yukawa couplings
$h_{fl}$ and $h_{bl}$, and d.o.f. $g_{fl}$ and $g_{bl}$,
respectively. These particles will be relatively light in the
low-temperature phase. We will consider also $g_{b}$ bosons and
$g_{f}$ fermions with variable couplings $h_{b}$ and $h_{f}$,
respectively. The values of the Yukawa couplings are constrained by
perturbativity of the theory, which sets a generic upper bound
$h_i\lesssim 3.5$ \cite{cmqw05}. In addition, we include $g_{l}$
light d.o.f., for which we  assume $h_i=0$. This model allows us to
explore several kinds of phase transitions.

\begin{figure}[htb]
\centering \epsfysize=6cm \leavevmode %
\epsfbox{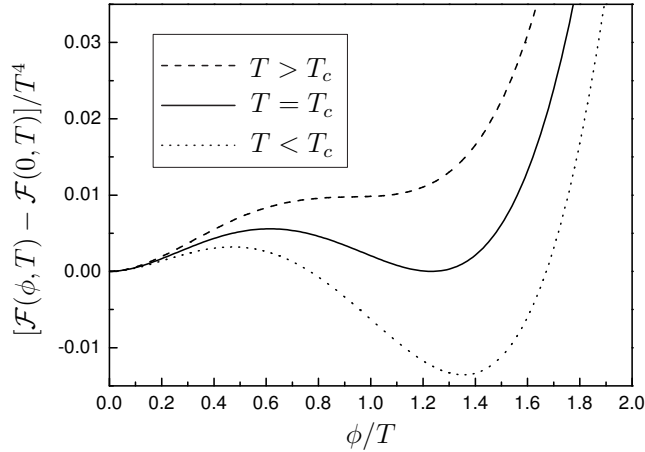} \caption{The free energy around $T=T_c$.}
\label{pot}
\end{figure}

For instance, choosing $v=246GeV$ we have a phase transition at the
electroweak scale. We obtain a good approximation for the free
energy of the Standard Model (SM) if we consider $g_{fl}=12$ fermion
d.o.f. with $h_{fl}\approx 0.7$ (corresponding to the top), and
$g_{bl}=6$ boson d.o.f. with $h_{bl}\approx 0.35$ (corresponding to
the transverse gauge vectors $W$ and $Z$). The rest of the SM d.o.f.
have $h\ll 1$, so their contribution to the $\phi$-dependent part of
the effective potential is negligible. They only contribute to $g_l$
in Eq. (\ref{ffinal}). To make the electroweak phase transition
strongly first-order we need to add some extra particles to the SM.
For our purposes, we don't need to refer to any specific extension
of the model. We choose $\lambda \approx 0.12$, which corresponds to
a Higgs mass $m_{H}=120GeV$, and we consider for the time being
adding equal numbers of bosons and fermions, with $g_{b}=g_{f}=10$
and $h_{b}=h_{f}=0.7$, which give a value $\phi _{c}/T_{c}\approx
1.3$ for the minimum of the potential at the critical temperature
(see Fig. \ref{pot}).

\begin{figure}[hbt]
\centering \epsfxsize=16cm \leavevmode \epsfbox{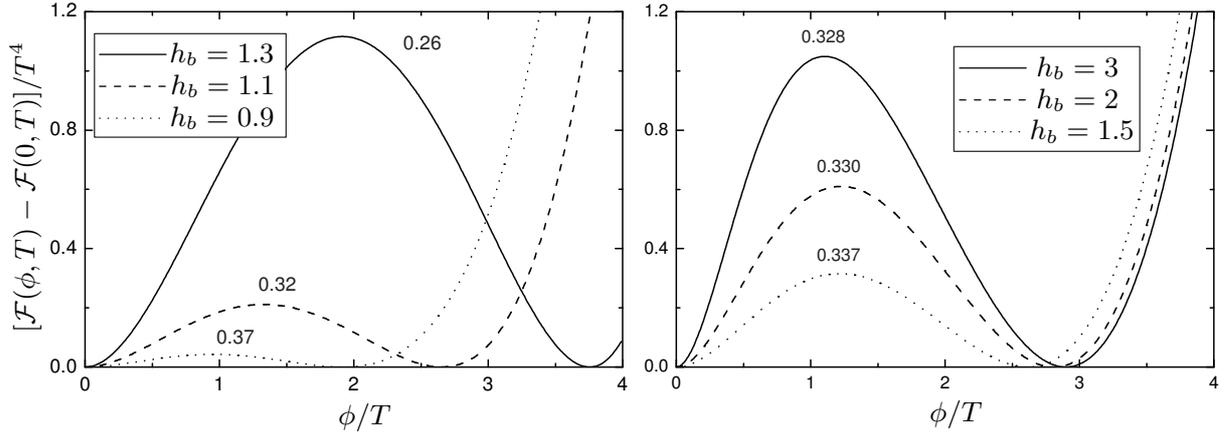}
\caption{The effective potential at $T=T_c$. Left: $h_f=0.7$. Right:
$h_f=h_b$. The numbers next to the curves indicate the corresponding
values of $T_c/v$.} \label{pothb}
\end{figure}

It is well known that heavy bosons enhance the strength of the phase
transition. We see in the left panel of  Fig. \ref{pothb} that the
minimum $\phi _{c}$, as well as the height of the barrier, increase
if we increase the value of $h_{b}$. Besides, the critical
temperature decreases. Indeed, for fixed $h_f$, according to Eq.
(\ref{t0}) the temperature $T_{0}$ vanishes for a value
$h_{b}=h_{b1}$ given by $g_{b}h_{b1}^{4}= 16\pi ^{2}\lambda +
g_{f}h_{f}^{4}+ g_{fl}h_{fl}^{4}- g_{bl}h_{bl}^{4}$. At this point,
a barrier appears in the zero-temperature effective potential, and
$\phi =0$ becomes a local minimum of $V(\phi)$. If $h_{b}$ is
increased further, the zero-temperature barrier increases as the
energy $\rho _{\Lambda }$ of the origin decreases. According to Eq.
(\ref{rholambda}), the two zero-temperature minima become degenerate
for a value $h_{b}=h_{b2}$ given by $g_{b} h_{b2}^{4}= 32\pi ^{2}
\lambda +g_{f}h_{f}^{4} +g_{fl}h_{fl}^{4}-g_{bl}h_{bl}^{4}$. For
this value of $h_b$ the critical temperature vanishes and
$\phi_c=v$. Beyond the value $h_{b2}$ there is no phase transition.

If we  keep $h_{f}=h_{b}$ as we increase $h_{b}$, the behavior is
quite different, since the fermions compensate the effect of the
bosons. As we can see in Fig. \ref{pothb} (right panel), the height
of the barrier increases more slowly with $h_b$, and the value of
$\phi _{c}/T$ does not change significantly. With this symmetric
choice of parameters, the false vacuum energy density
(\ref{rholambda}) does not depend on $h_{b}$ and $h_{f}$. Thus, the
origin will never be the stable minimum at $T=0$, and $ T_{c}$ will
never vanish. According to Eq. (\ref{t0}), the temperature $T_{0}$
does not vanish either, but it decreases as $1/h_{b}$. In the next
section we will analyze the effect of these two opposite variations.

For large couplings, the first-order phase transition becomes
stronger and the latent heat (\ref{lat1}) increases. The maximum
value $L/\rho _{R}=4/3$ will be achieved when {\em all} the
couplings $h_i$ are large. If $g_l\neq 0$, this maximum becomes
$4x/3$, with $x=(1-g_l/g_*)$. Consider the case $g_{bl}=g_{fl}=0$
and $ g_b=g_f=10$. For $h_f$ fixed, the maximum is reached at
$h_b=h_{b2}$, i.e., when $T_c=0$. For the case $h_f=0.7$ we obtain
the points in the $(R,r)$-plane that are shown in blue squares in
Fig. \ref{fconstr}. For $h_f=h_b$, on the contrary, there is no such
limit on $h_b$. In this case (black triangles in Fig.
\ref{fconstr}), as the coupling is increased the points accumulate
near the point $(x/3,4x/3)$, which is the corner of the
thermodynamically allowed region. In particular, for $g_l=0$, we see
that $L/\rho_R$ can be very close to the maximum $4/3$, even though
in this case $T_c$ does not vanish.

It is interesting to consider the case $h_f>h_b$. Notice, however,
that strong fermion couplings $h_{f}$ may destabilize the
zero-temperature potential, since they introduce negative quartic
terms in $ V\left( \phi \right) $. To stabilize the potential in the
case of a strongly coupled fermion, we can add a heavy boson with
the same coupling $h_f$ and d.o.f. $g_f$, and a mass
$m_b^2(\phi)=\mu_b^2+h_b^2\phi^2$ \cite{cmqw05}. If $\mu_b$ is large
enough, this boson will be decoupled from dynamics at $T\sim v$. The
maximum value of $\mu_b$ consistent with stability is obtained by
requiring the quartic term to be positive for $\phi\gg v$. It is
given by
\begin{equation}
\mu_b^2=h_f^2 v^2
\left[\exp\left(\frac{16\lambda\pi^2}{g_fh_f^4}\right) -1\right].
\label{mub}
\end{equation}
For a weakly-coupled fermion, $\mu_b$ is much larger than $v$ and
the stabilizing boson is completely decoupled. On the contrary, for
large $h_f$, $m_b$ approaches $m_f$ and we recover the previous
case. We have plotted in Fig. \ref{fconstr} the points of the
$(R,r)$-plane\footnote{The definitions of $\rho_{\Lambda}$ and
$\rho_R$ change slightly in this case.} corresponding to a variation
of $h_f$ (red circles). For small values of $h_f$  we have only the
fermion contribution, and the phase transition is weakly
first-order. In fact, there is a minimum value of $h_f$ for which
the phase transition becomes second-order. At this point, the latent
heat vanishes for a finite value of $\rho_{\Lambda}$. In contrast,
for large $h_f$ we have, as in the previous cases, a strongly
first-order phase transition.

As we see in Fig. \ref{fconstr}, in all the cases the total duration
of the phase transition becomes significant for large $h_i$. However,
the durations of supercooling and phase coexistence can be extremely
different in each case.

\section{The phase transition} \label{transi}

\subsection{Phase transition dynamics}

The nucleation and growth of bubbles in a first order phase
transition has been extensively studied (see e.g.
\cite{h95,qcd,hkllm93,ah92,dlhll92,eikr92}). According to the
conventional picture of bubble nucleation, at $T>T_c$ the field takes
the value $\phi=0$ throughout space. At $T<T_c$, bubbles of the
stable phase (i.e., with the value $\phi =\phi _{m}$ inside)
nucleate. We remark that in a weakly first-order phase transition
this picture may not work \cite{gkw91}. A quantitative determination
of the importance of subcritical bubbles requires in general
numerical calculations and is out of the scope of the present
investigation. For instance, lattice calculations for the case of the
{minimal} standard model (with unrealistically small values of the
Higgs mass) show that subcritical bubbles may play a significant role
at the onset of a weakly first-order electroweak phase transition
\cite{yy97}. Thus, our results for the amount of supercooling become
unreliable in the limit of very small values of the coupling $h_b$.

The thermal tunneling probability for bubble nucleation per unit
volume per unit time is \cite{a81,l83}
\begin{equation}
\Gamma \simeq A\left( T\right) e^{-S_{3}/T}.  \label{gamma}
\end{equation}%
The prefactor involves a determinant associated with the quantum
fluctuations around the instanton solution. In general it cannot be
evaluated analytically. However, the nucleation rate is dominated by
the exponential in (\ref{gamma}), so we will use the rough
estimation $A\left( T\right) \sim T_c^{4}$. The exponent in Eq.
(\ref{gamma}) is the three-dimensional instanton action
\begin{equation}
S_{3}=4\pi \int_{0}^{\infty }r^{2}dr\left[ \frac{1}{2}\left( \frac{d\phi }{dr%
}\right) ^{2}+ \Delta {\cal F}\left( \phi \left( r\right) ,T\right)
\right], \label{s3}
\end{equation}%
where $\Delta  {\cal F}\left( \phi ,T\right) = {\cal F}\left( \phi
,T\right)- {\cal F}\left( 0 ,T\right)$. The configuration $\phi
\left( r\right) $ of the nucleated bubble may be obtained by
extremizing this action. It obeys the equation
\begin{equation}
\frac{d^{2}\phi }{dr^{2}}+\frac{2}{r}\frac{d\phi
}{dr}=\frac{\partial {\cal F}}{\partial \phi} . \label{eqprofile}
\end{equation}%
Hence, $S_3$ coincides with the free energy that is needed to form a
bubble in unstable equilibrium between expansion and contraction. At
the critical temperature the bubble has infinite radius, so
$S_{3}=\infty $ and $\Gamma =0.$ In contrast, at $T=T_{0}$ the
radius vanishes, so $S_{3}=0$ and $\Gamma \sim T_{c}^{4}$, which is
an extremely large rate in comparison to $H^4\sim (T^2/M_P)^4$.
Therefore, the number of bubbles will become appreciable at a
temperature which is rather closer to $T_c$ than to $T_0$. Thus, in
order to have supercooling at $T\ll T_c$, the temperature $T_0$ must
not exist, so that the barrier between minima persists at $T= 0$.

After a bubble is formed, it grows due to the pressure difference at
its surface. There is a negligibly short acceleration stage until
the wall reaches a terminal velocity due to the viscosity of the
plasma (see, e.g., \cite{m00}). The velocity $v_{w}$ is determined
by the equilibrium between the pressure difference
$p_{-}-p_{+}=\mathcal{F}_{+}-\mathcal{F}_{-}\equiv\Delta
\mathcal{F}(T)$ and the force per unit area due to friction with the
surrounding particles, $f_{\rm friction}=\eta v_{w}$. Thus,
\begin{equation}
v_{w}(T)=\Delta \mathcal{F}(T)/\eta .  \label{vw}
\end{equation}%
The friction coefficient can be written as $\eta
=\tilde{\eta}T\sigma ,$ where $\tilde{\eta}$ is a dimensionless
damping coefficient that depends on the viscosity of the medium, and
$\sigma =\int \left( d\phi /dr\right)
^{2}dr $ is the bubble wall tension (for a review and a discussion see \cite%
{m04}).

We will assume that the system remains close to equilibrium, which
is correct if $v_{w}$ is small enough. If the wall velocity is lower
than the speed of sound in the relativistic plasma,
$c_{s}=\sqrt{1/3}$, the wall propagates as a deflagration front.
This means that a shock front precedes the wall, with a velocity
$v_{sh}>c_{s}$. For $v_{w}\ll c_{s}$, the latent heat is transmitted
away from the wall and quickly distributed throughout space. We can
take into account this effect by considering a
homogeneous reheating of the plasma during the expansion of bubbles \cite%
{h95,m01}. (For detailed treatments of hydrodynamics see, e.g., \cite%
{eikr92,hkllm93}).

The radius of a bubble that nucleates at time $t^{\prime }$ and
expands until  time $t$ is
\begin{equation}
R\left( t^{\prime },t\right) =R_{0}\left( T^{\prime }\right) \frac{a\left(
t\right) }{a\left( t^{\prime }\right) }+\int_{t^{\prime }}^{t}v_{w}\left(
T^{\prime \prime }\right) \frac{a\left( t\right) }{a\left( t^{\prime \prime
}\right) }dt^{\prime \prime }.  \label{radius}
\end{equation}%
The scale factor $a$ takes into account the fact that the radius of
a bubble increases due to the expansion of the Universe. The initial
radius $R_{0}$ can be calculated by solving Eq. (\ref{eqprofile})
for the bubble profile $\phi \left( r\right)$. It is roughly $\sim
T^{-1}$. Hence, $R_{0}$ can be neglected, since the second term in
Eq. (\ref{radius}), which is determined by the dynamics, depends on
the time scale $\delta t\sim H^{-1}\sim M_{P}/T^{2}.$

The fraction of volume occupied by bubbles is given by
\begin{equation}
f\left( t\right) =1-\exp \left\{ -\int_{t_{i}}^{t}\left( \frac{a\left(
t^{\prime }\right) }{a\left( t\right) }\right) ^{3}\Gamma \left( T^{\prime
}\right) \frac{4\pi }{3}R\left( t^{\prime },t\right) ^{3}dt^{\prime
}\right\} .  \label{fb}
\end{equation}%
The integral in the exponent gives the total volume of bubbles (in a
unit volume) at time $t$, ignoring overlapping. The complete
expression (\ref{fb}) takes into account bubble overlapping
\cite{gw81}. The factors of $a$ take into account that the number
density of nucleated bubbles decreases due to the expansion of the
Universe.

To integrate Eq. (\ref{fb}), we still need two equations in order to
relate the variables $T$, $a$ and $t$. Eqs (\ref{fcoex}) and
(\ref{s+}) give the relation \cite{m04}
\begin{equation}
T^{3}=\frac{-\Delta \mathcal{F}^{\prime }\left( T\right) }{2\pi
^{2}g_{\ast }/45}f+\frac{T_{c}^{3}a_{i}^{3}}{a^{3}},
\label{temperature}
\end{equation}%
where the first term, which is proportional to the released entropy
$\Delta s(T) f$, accounts for reheating, and the second term
accounts for the cooling of the Universe due to the adiabatic
expansion. Finally, the Friedmann equation (\ref{friedmann}) gives
the relation
\begin{equation} %
\frac{1}{a}\frac{da}{dt}=\sqrt{\frac{8\pi G}{3}\rho}, \label{dadt}
\end{equation}%
with $\rho =\rho _{+}\left( T\right) -\Delta \rho \left( T\right)
f$, where $\rho_+=\rho_{\Lambda}+g_*\pi^2T^4/30$, and $\Delta
\rho=-T\Delta{\cal F}'+ \Delta{\cal F}$.

The functions $\Delta {\cal F}(T)$ and $\Delta {\cal F}'(T)$ are
easily obtained by numerically finding the minimum $\phi_m(T)$. The
nucleation rate $\Gamma (T)$ can be calculated by solving
numerically Eq. (\ref{eqprofile}) for the bubble profile, then
integrating Eq. (\ref{s3}) for the bounce action, and using the
result in Eq. (\ref{gamma}). We solved Eq. (\ref{eqprofile})
iteratively by the overshoot-undershoot method\footnote{We have
checked our program by comparing with the results of Ref.
\cite{dlhll92} for the bounce and Ref. \cite{ma05} for the evolution
of the phase transition.}. The thermal integrals (\ref{integrals})
for the finite-temperature effective potential can be computed
numerically. However, we find that the computation time is lowered
significantly by using instead low-$x$ and high-$x$ expansions for
$I_{\pm}(x)$ (see the appendix).

\subsection{Numerical results}

\begin{figure}[ht]
\centering \epsfysize=6cm \leavevmode %
\epsfbox{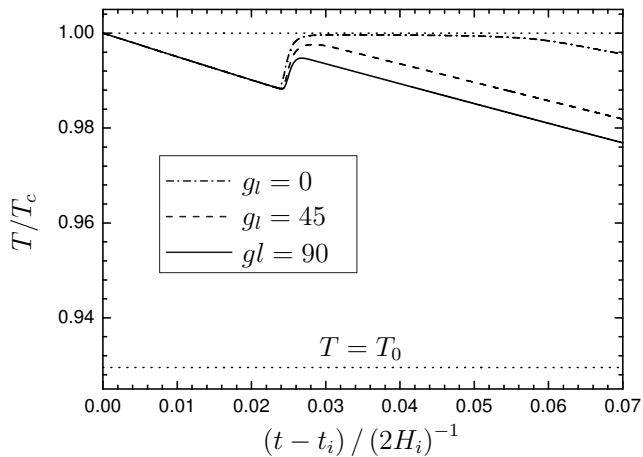} \caption{The temperature variation for the
potential of Fig. \ref{pot}.} \label{vargl}
\end{figure}

We begin by considering a phase transition at the electroweak scale,
with the free energy plotted in Fig. \ref{pot}. The development of
the phase transition depends on the specific heat of the thermal
bath, i.e., on the total number of d.o.f. We can take into account
the light d.o.f. of the SM by setting $g_l\approx 90$ in the last
term of Eq. (\ref{ffinal}). The friction coefficient $\tilde{\eta}$
depends on the model and its computation is not straightforward. For
the time being, let us assume $\tilde{\eta}\sim 1$. The solid curve
in Fig. \ref{vargl} shows the temperature variation during the phase
transition for this model. We observe a considerable reheating,
which indicates that the latent heat is comparable to the energy
density $\delta \rho_R $ needed to take the radiation back to
$T=T_c$. However, a phase coexistence stage is not achieved, which
reveals that $L\lesssim (4/3)\delta \rho_R$. Notice that the large
number of d.o.f ($g_{*}\sim 100$) makes the energy density of
radiation much larger than the latent heat. For lower values of
$g_{l}$, the thermal bath has a smaller specific heat and is more
easily reheated. This can be seen in the dashed and dashed-dotted
lines in Fig. \ref{vargl}.

We note that supercooling finishes at a temperature which is quite
closer to $T_{c}$ than to $T_{0}$. As mentioned before, this fact is
quite general, as it is due to the extremely rapid variation of the
nucleation rate, which becomes $\Gamma \sim T^{4}\gg H^{4}$ at
$T=T_{0}$. Notice also that the different curves in Fig. \ref{vargl}
coincide during supercooling. This is because in this stage the
relation between the dimensionless variables $T/T_{c}$ and $\tau
=\left( t-t_{i}\right) /\left( 2H_i\right) ^{-1}$ is almost
independent of any parameter of the model. Indeed, during
supercooling $T/T_{c}=a_i/a$, and the dependence of the scale factor
on time is given by $da/a=Hdt=(1/2)(H/H_{i})d\tau$. Since $a\approx
a_i$ and $H\approx H_{i} $ for $T\approx T_{c}$, we have
$d(T/T_{c})\approx -(1/2)d\tau$ as long as $T$ does not depart
significantly from $T_{c}$.

We can check the approximation (\ref{deltat}) for the total duration
of the phase transition. The relevant parameters $r=L/\rho_R$ and
$R=\rho _{\Lambda }/\rho _{R}$ are different for each curve in Fig.
\ref{vargl}, since $\rho _{R}$ depends on $g_{*}$. We obtain the
time lengths $\Delta t/\tilde{t} \approx 0.015$, $0.023,$ and
$0.055$. As expected, this approximation gives the correct value
only when $T$ gets close to $T_{c}$; otherwise, Eq. (\ref{deltat})
gives just a lower bound for $\Delta t$.

\begin{figure}[htb]
\centering
\epsfysize=6cm \leavevmode \epsfbox{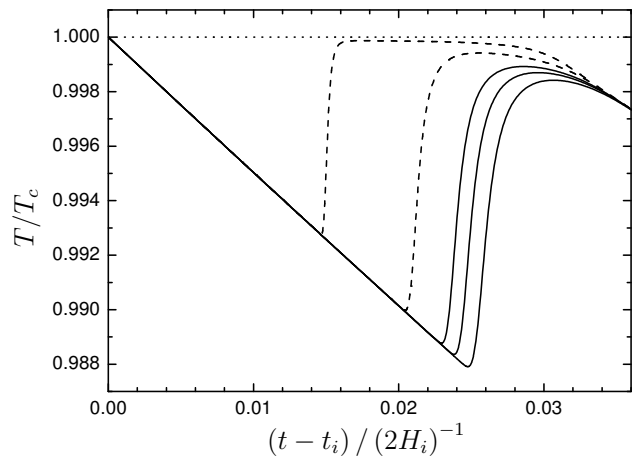} \caption{Temperature
variation for $g_l=30$. The three solid lines correspond to
$v=246GeV$ and, from right to left, to $\tilde{\eta}=50$,
$\tilde{\eta}=5$, and $\tilde{\eta}=0.5$. The dashed lines
correspond to $\tilde{\eta} =5$ and, from right to left, to $
v=100MeV$ and $v=10^{-3}eV$.} \label{vareta}
\end{figure}

Fixing now an intermediate value $g_{l}=30$ ($g_*\approx 65$), which
shows more clearly the effect of reheating, we consider three
different values of the friction (solid curves in Fig.
\ref{vareta}), in the range $\tilde{\eta}\sim 0.1 -  10$. These
correspond to velocities which have values between $v_{w}\sim 0.1$
and $v_{w}\sim 10^{-3}$ before reheating. We see that, as expected,
reheating begins earlier for larger initial velocities. However, a
variation of two orders of magnitude in $\tilde{\eta}$ does not
change significantly the amounts of supercooling and reheating. For
the rest of the paper, we will consider $\tilde{\eta}\sim 1$.

It is interesting to examine the role of the energy scale in the
dynamics of the transition. The model of Eq. (\ref{ffinal}) has a
single parameter with dimensions, namely, the minimum $v$, since the
masses of all the particles are of the form $m_i=h_iv$ (notice that
even the mass (\ref{mub}) of the stabilizing boson is proportional
to $v$). Thus, dimensionless quantities such as e.g. the ratio
$T_{c}/T_{0}$ will not be altered if we change the value of $v$.
This holds for all the quantities that are derived from the free
energy (e.g., $L/T^{4},\Gamma /T^{4}$), since the shape of the
normalized effective potential in Fig. \ref{pot} is unaffected.
Therefore, changing the scale $v$ will not affect the dynamics of
the transition, except for the expansion rate of the Universe Eq.
(\ref{dadt}), which depends on the ratio $M_{P}/T$.

To see the effect of such a change of scale, we have included in
Fig. \ref{vareta} a couple of examples in which the free energy is
the same as before, apart from the value of $v$. We considered the
QCD scale, $v\sim 100MeV$, and a scale $v\sim 10^{-3}eV$,
corresponding to a very recent phase transition (right and left
dashed lines, respectively). Again, the temperature decreases at the
same rate during supercooling, as explained above. However, bubble
nucleation and reheating begin sooner. This happens because at later
epochs the expansion rate $H$ is slower. As a consequence, the
nucleation rate $\Gamma$ becomes $\sim H^4$ with a smaller amount of
supercooling. In contrast, $L/T^{4}$ has the same value for any
scale $v$. Thus, since $\delta \rho_R$ is smaller, the temperature
gets closer to $T_{c}$. Hence, phase coexistence is favored in phase
transitions occurring at later times. The parameters $r$ and $R$
have the same values for all the curves in Fig. \ref{vareta}, and
Eq. (\ref{deltat}) yields $\Delta t/\tilde{t}\approx 0.029$.

\begin{figure}[htb]
\centering \epsfxsize=16cm \leavevmode \epsfbox{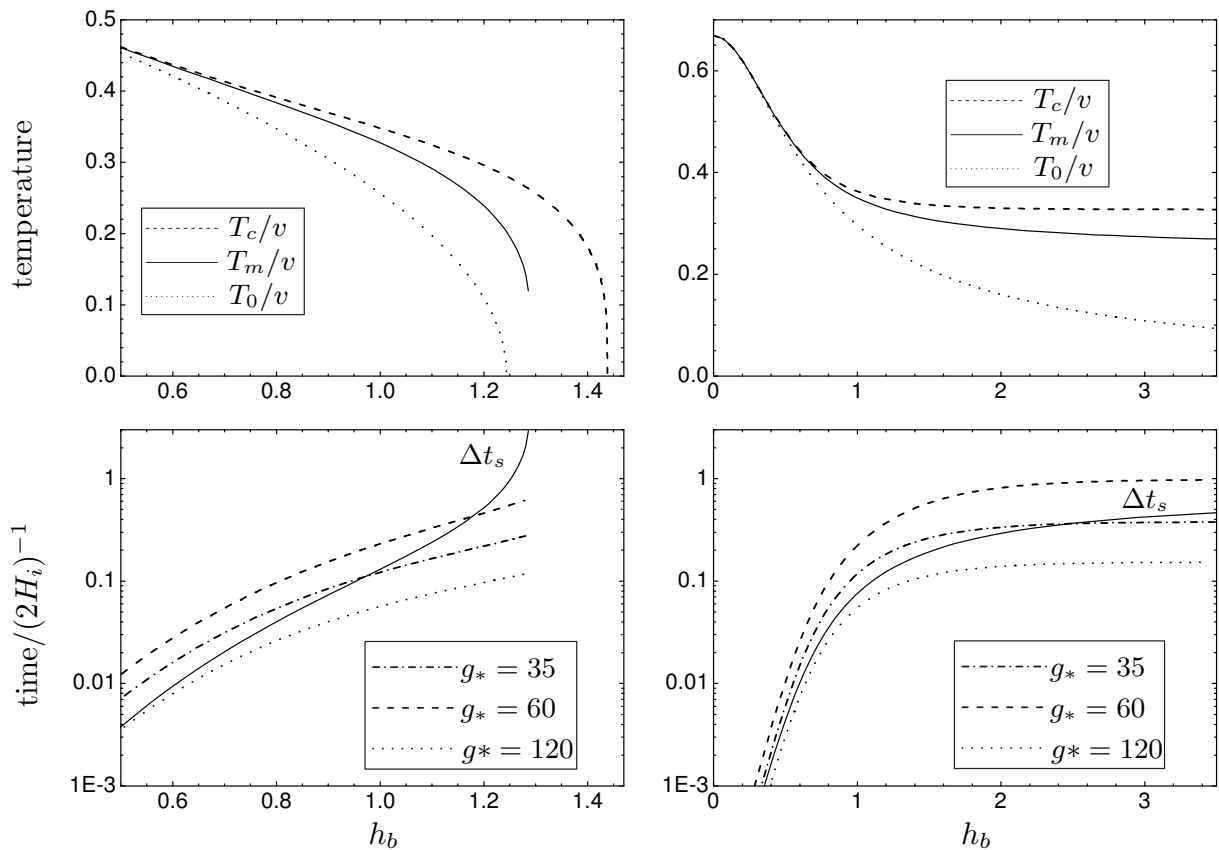}
\caption{Plots of the temperatures and time intervals as functions
of $h_b$, for $v=100MeV$ and $g_{f}=g_{b}=10$. Upper panels: The
temperatures $T_c$, $T_m$, and $T_0$ for $h_f=0.7$ (left) and
$h_f=h_b$ (right). Lower panels: The supercooling time $\Delta t_s$
corresponding to the upper panels (solid lines), and the total time
$\Delta t$ for different values of $g_l$.} \label{tmtiehb}
\end{figure}

So far we have varied the parameters $g_{l}$, $\eta $, and $v$,
which do not change the shape of the effective potential. We shall
now consider different values of the couplings $h_{b}$ and $h_{f}$,
fixing for simplicity $g_{b}=g_{f}=10$. We have checked that fixing
instead $ h_{b}$ and $h_{f}$ and considering different values of
$g_{b}$ and $g_{f}$ gives similar results. In what follows, we will
set $v=100MeV$. The result is shown in Fig. \ref{tmtiehb}. In the
upper panels we plot the temperature $T_{m}$ reached during
supercooling, together with the temperatures $T_{c}$ and $T_{0}$.
Notice that $T_m$ is always closer to $T_c$ than to $T_0$. The lower
panels show the time at which the temperature $T_{m}$ is reached,
i.e., the duration $\Delta t_{s}$ of supercooling (solid line). The
estimated duration $\Delta t$ of the phase transition is also shown,
for different values of $g_l$. As we increase the number of light
particles (without changing the potential), we obtain less reheating
for the same amount of supercooling. Hence, increasing $g_{l}$ gives
the same $\Delta t_{s}$ but a lower $\Delta t$.

The left panels of Fig. \ref{tmtiehb} illustrate the effect of a
variation of $h_b$ with $h_f$ fixed. As we have seen in the previous
section, in this case the temperature $T_{0}$ vanishes for a value
$h_{b}=h_{b1}$, where a zero-temperature barrier appears. For a value
$h_{b}=h_{b2}$, the critical temperature also vanishes. The
temperature $T_{m}$ lies between $ T_{0}$ and $T_{c}$, so it must
vanish for some value $h_{\max }$ with $h_{b1}<h_{\max }<h_{b2}$ (in
the present case, $h_{\max }\approx 1.3$). Our numerical calculation
does not allow us to plot the curve of $T_{m}$ up to this limit,
because the supercooling time diverges for $ h_{b}\rightarrow h_{\max
}$. This can be seen in the lower left panel. For $h_{b}>h_{\max }$
the system never gets out of the supercooling stage. On the contrary,
for $h_{b}<h_{\max }$ the phase transition completes in a finite
time. Regarding phase coexistence, it occurs when $\Delta t_{s}<
\Delta t$. For a given $g_l$, this happens up to a value of $h_{b}$
which is less than $h_{\max }$. Beyond that value, the supercooling
temperature $T_{m}$ is too low for the latent heat to provide the
required amount of reheating. Then, the estimation (\ref{deltat}) for
$\Delta t$ breaks down and the duration of the phase transition is
just given by $\Delta t\approx \Delta t_{s}$, since the phase
coexistence stage is replaced by a short reheating (see e.g. Fig.
\ref{vargl}).

If we now keep $h_{f}=h_{b}$ as we increase $h_{b}$ (right panels in
Fig. \ref{tmtiehb}), we see that $T_{c}$ does not vanish, and $T_{0}$
decreases like $1/h_{b}$ as expected. In this case, the supercooling
time $\Delta t_{s}$ does not diverge at any finite value of $h_{b}$,
and $\Delta t$ can be considerably larger than $\Delta t_s$. We see
that for small values of $g_{l}$ there is phase coexistence for any
value of $h_{b}$. On the contrary, for large values of $g_{l}$ there
is no phase coexistence at all, and the estimation for $\Delta t$
breaks down. The curves of $\Delta t$ saturate for $h_b$ large
because, for $g_l\neq 0$, the parameters $R$ and $r$ cannot get close
to their limits $R=1/3$, $r=4/3$.

\begin{figure}[htb]
\centering \epsfysize=6cm
\leavevmode \epsfbox{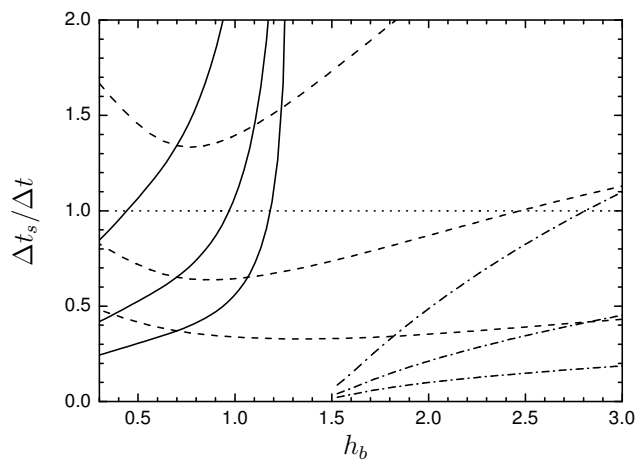} \caption{The fraction of time of
supercooling for the cases $h_f$ fixed (solid), $h_f=h_b$ (dashed),
and $h_f=h_b$ with $\mu_b\neq 0$ (dashed-dotted). For each set of
curves, from bottom to top $g_*=35$, $60$ and $120$.}
\label{tiempos}
\end{figure}

Let us now consider the case in which  $h_{f}=h_{b}$, but the boson
mass squared has a constant term $\mu_b^2$ given by Eq. (\ref{mub})
so it is partially decoupled from the thermodynamics. The curves we
obtained are similar to those in the right panels of Fig.
\ref{tmtiehb}, except that the temperatures meet at a finite value
$h_{\min}$. At this point the times fall to zero, since the phase
transition becomes second-order. We have plotted in Fig.
\ref{tiempos} the ratio $\Delta t_{s}/\Delta t$ for this case and
those of Fig. \ref{tmtiehb}. For each set of curves, the
supercooling fraction increases with $g_l$, since $\Delta t$
decreases. We also see that phase coexistence is favored for $
h_{f}\gtrsim h_{b}$. This is because fermions contribute to the
latent heat without enhancing the strength of the transition.

\begin{figure}[htb]
\centering \epsfysize=6cm \leavevmode
\epsfbox{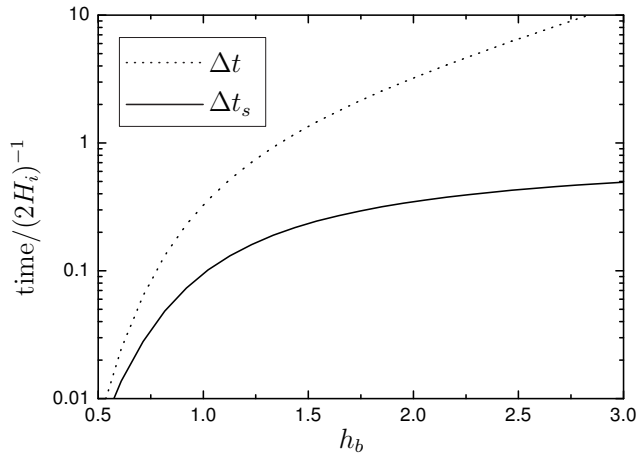} \caption{The time intervals for
$g_l=g_{bl}=g_{fl}=0$ and $h_f=h_b$.} \label{glcero}
\end{figure}

In general, the energy density of radiation, $\rho _{R}$, is much
larger than the latent heat, since only strongly coupled particles
contribute significantly to the latter. It is interesting to
consider the case in which there are no light d.o.f. at all, i.e.,
$g_{l}=g_{bl}=g_{fl}=0$. Only in this case the parameters can be
close to the thermodynamical limits $L\approx \rho _{R}+\rho
_{\Lambda }$, $\rho _{\Lambda }\approx \rho _{R}/3$. In the absence
of light particles, all the latent heat that is released during
bubble expansion is absorbed only by the heavy particles, which are
thus more easily reheated. Consequently, this scenario will be the
most favorable for phase coexistence.  We plot the time intervals in
Fig. \ref{glcero} for the case $h_f=h_b$. We find, as expected, that
the phase coexistence time is notably enhanced. For lower scales $v$
we will have the same $\Delta t$ but a smaller $\Delta t_s$.

\section{Phase transition dynamics and Cosmology \label{cosmo}}

The cosmological implications of a phase transition depend
drastically on the dynamics. In this section we discuss how the
different steps in the evolution, namely, supercooling, reheating and
phase coexistence, affect some of the observable products of a phase
transition.

\subsection{Late-time phase transitions and false vacuum energy}
\label{late}

Late-time phase transitions have been studied in connection to the
formation of large-scale structure and have been related, for
instance, to axions, domain walls and neutrino masses (see e.g.
\cite{fhw92}). In contrast to those occurring in the early Universe,
which take place in the presence of a hot plasma with a large number
of degrees of freedom, low-scale phase transitions happen in general
in a sector with a few d.o.f. and, consequently, a small specific
heat. Therefore, one expects a significant reheating during the phase
transition, and a long phase coexistence stage \cite{m06}. Indeed, we
have seen in section \ref{transi} that both a low $v$ and a small
$g_*$ favor a long phase coexistence period (see e.g. Figs
\ref{vargl} and \ref{vareta}). This stage can be significantly long
for strongly first-order phase transitions, as shown in Fig.
\ref{tmtiehb} (lower right panel). In particular, if all the
particles have strong couplings, Fig. \ref{glcero} shows that the
coexistence of phases can last for a time $\Delta t \gg t_i$.

Recently, late-time phase transitions have been considered with the
aim of (partially) solving the dark-energy problem. While the system
is trapped in the metastable phase, the energy density of the false
vacuum provides an effective cosmological constant. Thus, a false
vacuum energy $\rho_{\Lambda}\sim (10^{-3}eV)^4$ could explain the
observed acceleration of the Universe. This fact has motivated
several models in which a phase transition at a scale $v \sim
10^{-3}eV$ occurs in a hidden sector \cite{darke,m06,g00,chn04}. Such
a false vacuum must persist until the present epoch. Hence, since the
temperature of the hidden sector must be lower than that of photons,
$T_{\gamma}\sim 10^{-4}eV$, the system must be in the metastable
phase still at $T\ll v$ (for a discussion, see e.g., \cite{m06,g00}).
This could be achieved, in principle, in several ways, namely, due to
a low critical temperature $T_c \ll v$ \cite{chn04}, due to a large
amount of supercooling \cite{g00}, or due to a long phase coexistence
stage \cite{m06}.

The constraint on the temperature of the hidden sector,
$T<T_{\gamma}$, comes from the Big Bang Nucleosynthesis (BBN)
constraint on its radiation energy density, $\rho_R\lesssim
0.1\rho_{\gamma}$. Therefore, if the false vacuum energy
$\rho_{\Lambda}$ is to explain the observed dark energy, the
temperature of the system must be such that
\begin{equation}
\rho_R(T)\lesssim 10^{-5} \rho_{\Lambda}. \label{condlt}
\end{equation}
One possible way out of this limitation would be to assume that,
although the BBN constraint $T<T_{\gamma}$ was  satisfied for most
of the history of the Universe, when $T$ reached $T_c\sim 10^{-3}eV$
the system entered a long phase coexistence stage at constant
temperature \cite{m06}. Then, as $T_{\gamma}$ continued decreasing,
the temperature of the hidden sector was stuck at $T=T_c$. Thus, the
BBN condition $\rho_R<\rho_{\gamma}$ would be violated only at the
present epoch, avoiding the restriction (\ref{condlt}). As we have
seen in section \ref{termo}, a very long phase coexistence stage is
possible.  However, the effective cosmological constant during phase
coexistence is $\rho_{\Lambda}^{\rm eff}=\rho_{\Lambda}-\rho_R/3$
which, according to the thermodynamical bound Eq. (\ref{thconstr1}),
is negative and does not lead to accelerated expansion.

Due to  the  bound $\rho_R(T_c)>3\rho_{\Lambda}$, the condition
(\ref{condlt}) cannot be fulfilled  at $T\geq T_c$. This
automatically rules out any model in which false vacuum energy is
dominant because the phase transition is yet to occur. For instance,
in Ref. \cite{chn04} a potential with a negative quadratic term
$-m_{\phi}^2\phi^2$ is considered. In that model, the temperature is
assumed to be high enough that thermal corrections trap the system
in the false vacuum. Then, the condition (\ref{condlt}) is shown to
be achieved for a somewhat small value of a coupling constant
$\lambda$. Clearly the thermodynamical bound is strongly violated.
However, the thermal correction to the effective potential is
assumed to be $\sim T^2\phi^2$, which corresponds to keeping only
the quadratic term in the power expansion of the thermal integral
$I_-(x)$. It is then argued that the field is trapped at the origin
as long as $T^2$ is large enough to cancel the negative mass
squared. This would be correct in a second-order phase transition,
in which $T_c=T_0$. However, with the parameters of Ref.
\cite{chn04} the phase transition is strongly first-order. Hence, at
$T=T_0$ the field certainly lies in the minimum $\phi_m\neq 0$. The
critical temperature can in fact be much larger than $T_0$, as shown
in the upper-left panel of Fig. \ref{tmtiehb}, where we see that
$T_0$ can vanish while $T_c$ is still of order $v$.

Another possibility to attain condition (\ref{condlt}) is in a model
with a large amount of supercooling, so that $\rho_R(T_c)\sim
\rho_{\Lambda}$ but $T\ll T_c$. This is possible in a strongly
first-order phase transition. For the model considered in the left
panel of Fig. \ref{tmtiehb}, there is a maximum value of the
coupling $h_b=h_{\max}$ for which the supercooling temperature
$T_m\to 0$ and the duration of supercooling becomes infinite. It is
not clear, however, that the required amount of supercooling can be
achieved in a realistic model. Notice that, even when $T_0$ vanishes
(i.e., for $h_b=h_{b1}$), we still have $T_m\sim T_c$. In the
example of Fig. \ref{tmtiehb}, $\rho_R(T_m)\approx
2.37\rho_{\Lambda}$. Thus, it is necessary to go beyond
$h_b=h_{b1}$, i.e., to consider a model which has a barrier still at
$T=0$. In Ref. \cite{g00} the condition (\ref{condlt}) was
accomplished in a specific model with $T_0=0$ and some fine tuning
of the parameters. However, the thermal corrections were taken into
account only by introducing a term $\sim T^2\phi^2$. As pointed out
in Ref. \cite{m06}, this causes an unrealistically large value of
the latent heat, which violates the constraint Eq.
(\ref{thconstr2}).

\subsection{Electroweak baryogenesis and baryon inhomogeneities}

It is well known that the electroweak phase transition could be the
framework for the generation of the baryon number asymmetry of the
Universe (BAU). A first-order electroweak phase transition provides
the three Sakharov's conditions for the generation of a BAU,
although physics beyond the minimal Standard Model is mandatory in
order to obtain a quantitatively satisfactory result \cite{ckn93}.
Due to $CP$ violating interactions of particles with the bubble
walls, a net baryon number density $n_B$ is generated around the
walls of expanding bubbles. Assuming that $CP$ violation is strong
enough and that the baryon number violating sphaleron processes are
suppressed in the broken symmetry phase, the resulting $n_B$ depends
on the bubble wall velocity $v_w$. If the velocity is too large,
sphalerons will not have enough time to produce baryons. On the
other hand, for very small velocities thermal equilibrium is
restored and sphalerons erase any generated baryon asymmetry. As a
consequence, the generated baryon number has a peak at a given wall
velocity, which is generally $v_w\sim 10^{-2}$
\cite{lmt92,ckn92,ck00}.

As we have seen, reheating is always appreciable, even if there is
no phase coexistence\footnote{An exception could be the case of an
extremely supercooled electroweak phase transition, for which
reheating may be negligible. Such a model has been considered
recently in Ref. \cite{nqw07}.}. The temperature rise causes the
wall velocity to descend significantly. Thus, baryogenesis is either
enhanced or suppressed, depending on which side of the peak of
$n_B(v_w)$ the initial velocity lies \cite{h95,m01}. Furthermore,
baryon inhomogeneities arise due to the variation of $v_w$.
Electroweak baryon inhomogeneities may survive until the QCD scale
\cite{s03,jf94} and affect the dynamics of the quark-hadron phase
transition \cite{s03,cm96,h83}. The geometry of the inhomogeneities
was studied in Refs. \cite{ma05,h95}. Since baryon number is
generated near the bubble walls, a spherical inhomogeneity with a
radial profile is formed inside each expanding bubble.

Notice that bubble nucleation stops as soon as reheating begins. In
fact, due to the exponential variation of the nucleation rate with
temperature, most bubbles are formed in a small interval $\delta
t_{\Gamma}$ around the time $t_m$ at which the minimum temperature
$T_m$ is reached \cite{ma05}. This interval is in general much
shorter than the time it takes expanding bubbles to complete the
phase transition. Therefore, it is a good approximation to assume
that all bubbles are created at $t=t_m$. At a later time, their
walls are moving with  a velocity $v_w(T(t))$. Hence, all the
inhomogeneities have the same profile.

In Refs. \cite{ma05,h95}, the size and amplitude of the electroweak
baryon inhomogeneities were investigated using a simple effective
potential, whose parameters were adjusted so as to give the desired
values of the thermodynamic parameters. This approximation allows to
vary independently parameters such as, e.g., the latent heat or the
bubble-wall tension. These parameters, though, are generally related
in a non-trivial way, which depends on the extension of the SM that
is considered. For instance, a strongly first-order phase transition
will have in general a considerable amount of supercooling, and also
a large latent heat. However, the relative importance of
supercooling  and reheating depends significantly on the specific
model, as can be seen, for instance, in Fig. \ref{tiempos}.

The amplitude of the baryon inhomogeneities, $n_{B\max}/n_{B\min}$,
is bounded by the ratio of the highest and lowest wall velocities
reached during bubble expansion, $v_{\max}/v_{\min}$. If $L$ is
freely varied, one can achieve values $v_{\max}/v_{\min} \sim 100$
or higher \cite{ma05,h95}. However, in a specific extension of the
SM this will not be necessarily so. To examine a more realistic
situation, we have considered extensions of the SM as in the
previous sections. We find that, if we add a strongly coupled boson,
or a boson and a fermion with $h_f\leq h_b$, the velocity variation
is in general $v_{\max}/v_{\min}\sim 1$. We find a sizeable ratio
only in the case in which the fermion dominates. The addition of
strongly coupled fermions was investigated in Ref. \cite{cmqw05}, in
order to make the electroweak phase transition strongly first-order.

\begin{figure}[htb]
\centering \epsfysize=6cm \leavevmode
\epsfbox{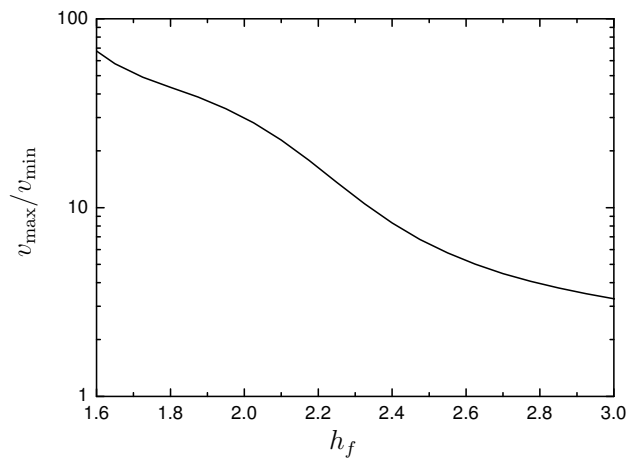} \caption{The ratio $v_{\max}/v_{\min}$ as a
function of $h_f$.} \label{veloc}
\end{figure}

\begin{figure}[htb]
\centering \epsfysize=6cm \leavevmode \epsfbox{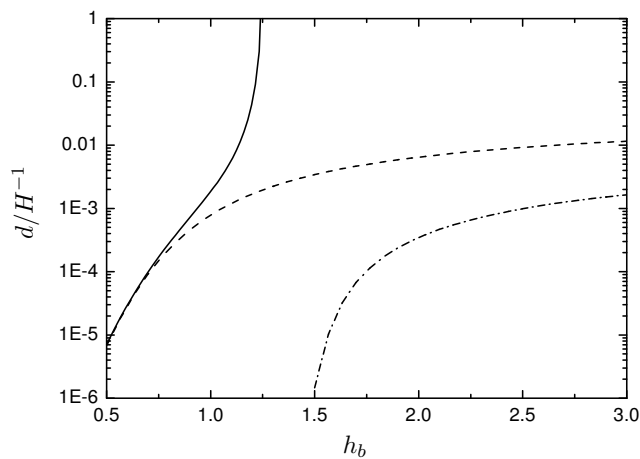}
\caption{The distance $d$ between centers of nucleation for $g_f=0$
(solid line), $g_f=g_b=10$ and $h_f=h_b$ (dashed line), and
$g_f=g_b=10$, $h_f=h_b$, with $\mu_b\neq 0$ (dashed-dotted).}
\label{dist}
\end{figure}

Let us consider for simplicity an extension with $g_f=10$ fermionic
d.o.f. with mass $m_f=h_f\phi$, and stabilizing bosons with
$g_b=g_f$, $h_b=h_f$, and a dispersion relation
$m_b^2=\mu_b^2+h_b^2\phi^2$, with $\mu_b$ given by Eq. (\ref{mub}).
We obtain the plot of Fig. \ref{veloc}. For $h_f$ in the range of
the figure the value of the order parameter is $\phi_c/T_c>1$, as
required by electroweak baryogenesis.
The distance scale of the inhomogeneities is given by the final size
of bubbles, which depends on the distance between centers of
nucleation. Thus, it can be roughly estimated as $d\sim n^{-1/3}$.
For the present case we obtain the dashed-dotted curve in Fig.
\ref{dist}. Our results for the distance $d$ agree in order of
magnitude with those of Refs. \cite{ma05,h95}. However, we see that
the amplitude of the inhomogeneities can be important only for small
values of $h_f$. In particular, $v_{\max}/v_{\min}\gtrsim 100$ is
reached for values of $h_f$ for which $\phi_c/T_c<1$. Therefore,
baryon inhomogeneities of significant amplitude are not likely
produced in the electroweak phase transition.

\subsection{Topological defects and magnetic fields}

If a global $U\left( 1\right) $ symmetry is spontaneously broken at
a first-order phase transition, the phase angle $\theta $ of the
Higgs field takes different and uncorrelated values inside each
nucleated bubble. When bubbles collide, the variation of the phase
from one domain to another is smoothed out. According to the {\em
geodesic rule}, the shortest path between the two phases is chosen
\cite{k76}. When three bubbles meet, a vortex (in two spatial
dimensions) or a string (in 3d) may be trapped between them. This
mechanism can be generalized to higher symmetry groups and other
kinds of topological defects.

If the dynamics for the phase  $\theta $ is not taken into account,
the number density of defects depends only on the final bubble size.
The probability of trapping a string at the meeting point of three
bubbles is $1/4$. Thus, the string density (length per unit volume)
is $\sim 1/4d^2$, where $d$ is the distance between bubble centers
\cite{kv95}. Fig. \ref{dist} shows the different possibilities for
the length $d$. For stronger phase transitions, the bubble separation
is larger, since the nucleation rate is more suppressed.

Taking into account the dynamics of phase equilibration, the number
density of defects depends also on the velocity of bubble expansion.
If the latter is much less than the velocity of light, the
equilibration between the phases of two bubbles may complete before
a third bubble meets them, thus reducing the chances of trapping a
string. Consequently, reheating hinders the formation of topological
defects.

In the case of a gauge theory,
a spatial variation of the phase $\theta$ is linked to a variation
of the gauge field \cite{kv95}. As a consequence, a magnetic field
is generated together with the phase difference in the collision of
two bubbles. Then, one can say that a vortex is formed whenever a
quantum of magnetic flux is trapped in the unbroken-symmetry region
between three bubbles. The phase equilibration process is thus
related to flux spreading, and depends on the conductivity of the
plasma. Bubble collision constitutes also a mechanism for generating
the cosmic magnetic fields (see e.g. \cite{gr98}). This mechanism
may take place at the electroweak phase transition, where unstable
cosmic strings and hypermagnetic fields may be formed. The latter
are subsequently converted to $U\left( 1\right) _{\rm em}$ magnetic
fields.

A detailed calculation of the density of defects and the magnitude
of the magnetic fields is beyond the scope of this paper and we
leave it for future research. Although some simulations have been
made (see, e.g., \cite{bkvv95}), several simplifications are
generally used, which include assuming a constant nucleation rate
and a constant bubble wall velocity. As we have seen, this situation
is hardly realistic. Moreover, the formation of topological defects
and magnetic fields depend strongly on the dynamics of the phase
transition. In particular, a long phase coexistence stage with a
very slow bubble expansion will affect significantly the mechanism
of phase equilibration during bubble percolation.

\section{Conclusions} \label{discus}

In this article we have investigated the different stages in the
development of first-order phase transitions of the Universe. In
particular, we have studied the amounts of supercooling and
reheating. If the entropy discontinuity $\Delta s (T_c)$ is larger
than the entropy decrease $\delta s=s(T_c)-s(T_N)$ during
supercooling, a phase-coexistence stage is reached. Then, the total
duration of the phase transition can be calculated analytically. The
ratio $\Delta t/ (2H)^{-1}$ depends only on the parameters
$r=L/\rho_R$ and $R=\rho_{\Lambda}/\rho_R$. If $\Delta s (T_c)\leq
\delta s$, supercooling lasts for a time which is longer than
$\Delta t$. In this case, there is no phase coexistence, and $\Delta
t$ gives only a lower bound for the total duration of the phase
transition. We have shown that thermodynamics constrain these
parameters to the region $R\leq 1/3$, $r\leq R+1$. These constraints
should be taken into account when the dynamics of a particular phase
transition is considered, since approximations for the effective
potential may violate them, and thus the analysis may lead to
incorrect results.

With the help of a simple model, we have analyzed numerically the
role of different parameters in the dynamics of the phase
transition. We have verified that phase coexistence is more likely
in later phase transitions, since both a lower energy scale and a
smaller number of degrees of freedom favor reheating. In addition,
we have seen that changing the viscosity of the surrounding medium
does not affect significantly the dynamics of supercooling and
reheating, although it affects the velocity of bubble walls. The
incorporation of bosons to a given model strengthens the phase
transition, so the effect on the dynamics is to enlarge the latent
heat and suppress the nucleation rate. As we have seen, the latter
effect is in general stronger, so adding bosons favors supercooling.
On the contrary, adding fermions in general weakens the phase
transition and at the same time increases the number of d.o.f. We
have checked that in this case phase coexistence is favored.

We have studied how our general results on phase transition dynamics
may affect some of the cosmological consequences. For instance, in
the case of dark energy from a phase transition, we have shown that
the thermodynamical bounds rule out some models. Besides, we have
analyzed the effect of dynamics on two important parameters, namely,
the number density of bubbles and the amplitude of the velocity
variation during reheating. As we have seen, these quantities are
relevant for the generation of different cosmological relics, e.g.,
baryon inhomogeneities, topological defects and magnetic fields. In
particular, we have found that it is difficult to obtain baryon
inhomogeneities of sizeable amplitude in realistic models of the
electroweak phase transition.

We believe that our results on the dynamics can be applied to a wide
class of phase transitions of the Universe, and the discussion on
the cosmological consequences can be extended to several interesting
possibilities, such as, e.g.,  the formation of baryon
inhomogeneities in the quark-hadron phase transition \cite{w84} or
the generation of gravitational waves \cite{gs07}.

\acknowledgements

This work was supported in part by Universidad Nacional de Mar del
Plata, Argentina, grants EXA 338/06 and 365/07. The work by A.D.S.
was supported by CONICET through project PIP 5072. The work by A.M.
was supported by FONCyT grant PICT 33635.

\appendix

\section{Approximations for the thermal integrals} \label{apen}

In this appendix we consider expansions of the functions
$I_{\pm}(x)$ for small $x$ and large $x$. The integrals in Eq.
(\ref{integrals}) can be evaluated numerically. However, a numerical
computation in the effective potential increases significantly the
total computation time. Indeed, notice that for each temperature, we
must find the minimum $\phi_m(T)$ to compute several quantities
derived from ${\cal F}(T)$. Moreover, the calculation of the bounce
action $S_3(T)$ requires the time-demanding overshoot-undershoot
technique to solve Eq. (\ref{eqprofile}) for the bubble profile at
each $T$. Therefore, it is useful to employ analytical
approximations for the thermal integrals.

Following the derivation of Ref. \cite{dj74}, we can obtain the
expansions of  $I_{\pm }\left( x\right) $ in powers of $x.$ For
bosons we have
\begin{eqnarray}
I_{-}\left( x\right) &=&-\frac{\pi ^{4}}{45}+\frac{\pi ^{2}}{12}x^{2}-\frac{%
\pi }{6}x^{3}-\frac{x^{4}}{32}\log \frac{x^{2}}{a_{b}} \\
&&-2\pi ^{7/2}\sum_{l=1}^{\infty }\left( -1\right) ^{l}\frac{\zeta \left(
2l+1\right) }{\left( l+2\right) !}\Gamma \left( l+\frac{1}{2}\right) \left(
\frac{x}{2\pi }\right) ^{2l+4},  \nonumber
\end{eqnarray}%
where $a_{b}$ is given by $\log a_{b}=3/2-2\gamma +2\log \left( 4\pi
\right) ,$ with $\gamma $ the Euler constant; $\zeta $ is the
Riemann zeta function,
and $\Gamma $ is the Gamma function. The expansion for fermions is%
\begin{eqnarray}
I_{+}\left( x\right) &=&-\frac{7\pi ^{4}}{360}+\frac{\pi ^{2}}{24}x^{2}+%
\frac{x^{4}}{32}\log \frac{x^{2}}{a_{f}} \\
&&+\frac{\pi ^{7/2}}{4}\sum_{l=1}^{\infty }\left( -1\right) ^{l}\frac{\zeta
\left( 2l+1\right) }{\left( l+2\right) !}\left( 1-\frac{1}{2^{2l+1}}\right)
\Gamma \left( l+\frac{1}{2}\right) \left( \frac{x}{\pi }\right) ^{2l+4},
\nonumber
\end{eqnarray}%
where $a_{f}$ is given by $\log a_{f}=3/2-2\gamma +2\log \pi $. For any
value of $x$ we can get the desired precision by keeping enough terms in
these expansions. For example, keeping up to $l=5$ in $I_{-}$ and $l=12$ in $%
I_{+},$ we obtain a precision of $10^{-8}$ for $0\leq x\leq 2$.

The expansion for large $x$ can be obtained by changing the variable
of integration to $z=\sqrt{y^{2}+x^{2}}$ and expanding the logarithm
in Eq. (\ref{integrals}) in powers of $e^{-z}$ (see Ref.
\cite{ah92}),
\begin{equation}
I_{\mp }\left( x\right) =-\sum_{k=1}^{\infty }\frac{\left( \pm 1\right)
^{k+1}}{k}\int_{x}^{\infty }dz\ z\sqrt{z^{2}-x^{2}}e^{-kz}.  \label{intxgran}
\end{equation}%
For each $k,$ the integral yields $x^{2}K_{2}\left( kx\right) /k$, where $%
K_{2}$ is the $n=2$ modified Bessel function of the second kind \cite%
{abramowitz} $K_{n}\left( z\right) $. Hence, we obtain the expansions
\begin{equation}
I_{\mp }=-x^{2}\sum_{k=1}^{\infty }\frac{\left( \pm 1\right) ^{k+1}}{k^{2}}%
K_{2}\left( kx\right) .  \label{ibessel}
\end{equation}%
Notice that the integrals in Eq. (\ref{intxgran}) are of the order of $%
e^{-kx}$, so the terms in this expansion decrease with powers of
$e^{-x}$. Therefore, in general we will obtain the desired precision
by considering a few terms. For example, for $x\geq 10$ we obtain
$\Delta I/I\lesssim 10^{-10} $ by keeping only the first two terms
in (\ref{ibessel}). For $x\geq 2$, keeping terms up to $k=7$ in the
expansion gives a precision $\Delta I/I\lesssim 10^{-8}$. As a rough
estimation of the error of the truncated expansion, we note that the
$k$-th term is $\sim x^{2}e^{-kx}/k^{2},$ and the error is given by
the ratio of the $\left( k+1\right) $-th term to the first term,
$\Delta I/I\sim e^{-kx}/k^{2}$.

\end{document}